\documentclass
[preprint,prb,a4paper,amsfonts,amssymb,floats,titlepage,fleqn,footinbib,lengthcheck,twocolumn,tightenlines,balancelastpage,10pt,unsortedaddress,showpacs]{revtex4}%
\usepackage{natbib}
\usepackage{amssymb}
\usepackage{amsfonts}
\usepackage{amsmath}
\usepackage{graphicx}
\usepackage{color}
\usepackage{longtable}
\usepackage{revsymb}
\usepackage{units}
\usepackage{nicefrac}
\usepackage{textcomp}
\usepackage{epsfig}
\usepackage{footmisc}
\usepackage{soul}%
\begin{document}
\title{Evolution of the magnetism of Tb(Co$_{x}$Ni$_{1-x}$)$_{2}$B$_{2}$C}
\author{M. ElMassalami,$^{1}$ H. Takeya,$^{2}$ A. M. Gomes,$^{1}$ T. Paiva,$^{1}$ and
R. R. \surname{dos Santos}$^{1}$}
\affiliation{$^{1}$Instituto de Fisica, Universidade Federal do Rio de Janeiro, Caixa
Postal 68528, 21941-972 Rio de Janeiro, Brazil,}
\affiliation{$^{2}$National Institute for Materials Science,1-2-1 Sengen, Tsukuba, Ibaraki
305-0047, Japan,}

\begin{abstract}
The magnetic properties of polycrystalline Tb(Co$_{x}$Ni$_{1-x}$)$_{2}$B$_{2}%
$C $(x=0.2,0.4,0.6,0.8)$ samples were probed by magnetization, specific heat,
$ac$ susceptibility, and resistivity techniques. For $x\neq0.4$, the obtained
curves are consistent with the features expected for the corresponding
magnetic modes, namely $\vec{k}_{1}=(0.55,0,0)$ at $x=0$; $\vec{k}%
_{2}=(\nicefrac{1}{2},0,\nicefrac{1}{2})$ at $x=$ 0.2; $\vec{k}_{3}%
=(0,0,\nicefrac{1}{3})$ at $x=$ 0.6, and $\vec{k}_{4}=(0,0,0)$ at $x=$ 0.8 and
1. For $x=0.4$, even though the neutron diffraction indicates a $\vec{k}_{2}$
mode, but with a reduced magnetic moment, the magnetization, the $ac$
susceptibility, and resistivity indicate two magnetic events; furthermore,
deviation from Curie-Weiss behavior is observed below 150 K for this sample.
These features, together with the evolution of both magnetic moment and
critical temperature, are attributed to an interplay between competing
magnetic couplings; for the particular $x=0.4$ case, additional factors such
as crystalline electric field effects may be in operation.

\end{abstract}
\date{Version 6 -- \today}

\maketitle

\section{Introduction}

The crystal structure of various families of compounds consists of magnetic
layers that are separated by nonmagnetic spacers. As in most cases, the
interlayer couplings in such arrangement are taken to be mediated by extended
electronic orbitals; therefore, a variation in electronic parameters of the
spacers is expected to have a profound influence on the magnetism, as well as
on the overall physical properties of the whole system. Such influence is most
dramatically manifested whenever the energy of the magnetic coupling is close
to that of the electronic interactions such as pairing or electron-phonon
interactions. Indeed, striking manifestations are evident in the magnetic and
transport properties of layered materials such as magnetic multilayers,
high-$T_{c}$ cuprates, pnictides superconductors, intermetallic magnetic
superconductors, and heavy fermion compounds.

The intermetallic borocarbides $RT_{2}$\textrm{B}$_{2}$\textrm{C} ($R$ is a
rare earth atom with 4$f$ moment, and $T$ is an unpolarized transition metal
atom) comprise one of these magnetically layered families wherein the magnetic
$R$C are stacked on the nonmagnetic $T_{2}$B$_{2}$ spacers, forming an $\cdots
R$C-$T_{2}$B$_{2}$-$R$C$\cdots$
pile.\cite{Muller01-interplay-review,Gupta06-Review-Borocarbides} The
electronic properties of the $T_{2}$\textrm{B}$_{2}$ spacers are crucial to
the determination of the overall physical properties, as manifested by the
surge of a wealth of exotic phenomena; see
Refs.{\onlinecite{Muller01-interplay-review,Gupta06-Review-Borocarbides,Cho96-RNi2B2C-deGeness,Canfield-RNi2B2C-Hc2-review}}%
, and references therein. Of particular interest to the present study is the
influence that the electronic properties exercise on the stabilization of the
ground-state magnetic structure. This is best illustrated by how the magnetic
arrangement of the Tb sublattice for different \textrm{Tb(Co}$_{x}$%
\textrm{Ni}$_{1-x}$\textrm{)}$_{2}$\textrm{B}$_{2}$\textrm{C} compositions is
controlled by the relative abundance of Ni and Co, $x$%
:\cite{Tb(CoNi)2B2C-Nd-2012} the longitudinal spin density wave, LSDW,
$\vec{k}_{1}=(0.55,0,0)$ state at $x=0$ is transformed into a collinear
$\vec{k}_{2}=$($\nicefrac{1}{2},0,\nicefrac{1}{2}$) antiferromagnetic,
AFM,\ state at $x=$ 0.2, 0.4; then into a transverse $c$-axis modulated
$\vec{k}_{3}=$($0,0,\nicefrac{1}{3}$) mode at $x=$ 0.6, and finally into a
simple ferromagnetic, FM, structure, $\vec{k}_{4}=(0,0,0)$, at $x=$ 0.8 and
1.0 (see Table \ref{TabI-Tc-Mag}).

This tuning in of magnetic modes by a simple variation of $x$ highlights the
importance of electronic band structures (which are dominated by the spacer 3d
orbitals):\cite{Pickett94-electronic-structure,Matthias94-electronic-structure,Ravindran98-Y(NiCo)2B2C,Lee94-electronic-structure,Coehoorn94-RNi2B2C-electronic-structure}
indeed, the generalized susceptibility calculations of Rhee \textit{et
al.}\cite{Rhee95-generalized-susc} predict a variety of magnetic modes, one of
which was verified only recently.\cite{Tb(CoNi)2B2C-Nd-2012} An alternative
theoretical analysis was proposed by Bertussi \textit{et al.}%
,\cite{Bertussi09-U-J-PhaseDiagram} based on an effective microscopic model in
which the interplay between local moments ordering and superconductivity is
explicitly taken into account. This interplay, as well as the specific
layering character of borocarbides\cite{03-Layer-Tc-borocarbides} have been
tested in one spatial dimension (1D), which is amenable to unbiased
calculations, and, as it turned out, the key element consists of conduction
electrons being mediators of both pairing and magnetic coupling. In this way,
a phase diagram relating the magnetic coupling to a succession of magnetic
modes is established,\cite{Bertussi09-U-J-PhaseDiagram} and this phase diagram
is found to be in fair agreement with the experimental
observations.\cite{Tb(CoNi)2B2C-Nd-2012}%

\begin{table*}[tbp] \centering
\caption{Selected magnetic properties of Tb(Co$_{x}$Ni$_{1-x}$)$_{2}$B$_{2}$C (adapted from Ref.\ \onlinecite{Tb(CoNi)2B2C-Nd-2012}). $\mu_{\rm ND}$ ($\mu_{\rm M}$) is the zero-field (90 kOe) value obtained from neutron diffraction (magnetization isotherm). $T_{cr}(x)$ is the critical point which is taken from each experimental curve.  As seen, except for $x=0.4$, the obtained transition point decreases monotonically with $x$, and, interestingly, tracks the behavior of the unit-cell volume;\cite {Tb(CoNi)2B2C-Nd-2012} this emphasizes the dependence of both on the electronic structure. \color{black} }%

\begin{tabular}
[c]{ccccccc}\hline\hline
$x$ & 0\footnotemark[1] & 0.2 & 0.4 & 0.6 & 0.8 & 1.0\footnotemark[2]\\\hline
\multicolumn{1}{l}{structure} & LSDW & AFM & AFM & TSDW & FM & FM\\
\multicolumn{1}{l}{$\overrightarrow{k}$} & ($0.55,0,0$) &
($\nicefrac{1}{2},0,\nicefrac{1}{2}$) & ($\nicefrac{1}{2},0,\nicefrac{1}{2}$%
) & ($0,0,\nicefrac{1}{3}$) & ($0,0,0$) & ($0,0,0$)\\
\multicolumn{1}{l}{$\left\vert \vec{\mu}\right\vert _{\text{ND}}$ ($\mu_{B}$)}
& 7.8 & 7.6(1) & 3.7(2) & 8.5(2) & 8.7(2) & 7.6\\
\multicolumn{1}{l}{$\left\vert \vec{\mu}\right\vert _{\text{M(90 kOe)}}$
($\mu_{B}$)} & 7.4(1) & 4.1(1) & 7.6(2) & 7.7(1) & 7.6(1) & 7.2\\
$T_{cr}$ (K) from \ $C(T,x)$ & 15\footnotemark[3] & 10.2(2) & $T_{2}=4.8(3)$
K\footnotemark[4] & 7.6(3) & 5.9(2) & 6.6\footnotemark[5]\\
$"\ \ "\ \ \ \ \ \ \ \ \ \chi_{dc}(T,x)$ & - & 10.4(2) & $T_{1}=11.0(2)$ K,
$T_{2}=4.3(3)$ K & 7.6(2) & 5.9(2) & -\\
$"\ \ "\ \ \ \ \ \ \ \ \chi_{ac}(T,x)$ & - & 10.3(2) & $T_{1}=11.8(2)$ K,
$T_{2}=4.0(2)$ K) & 7.6(2) & 5.9(2) & -\\\hline\hline
\end{tabular}
\footnotetext[1]{Ref.{\onlinecite{Lynn97-RNi2B2C-ND-mag-crys-structure}}%
}\footnotetext[2]{Ref.{\onlinecite{09-MS-TbCo2B2C}.}}\footnotetext[3]{see
Ref.{\onlinecite{Lynn97-RNi2B2C-ND-mag-crys-structure}. The specific heat of
TbNi}$_{{2}}${B}$_{{2}}${C\ from the present batch peaks at }$T_{p}=13.2(2)$
K{\ }}\footnotetext[4]{Only one event can be discerned [see text and
Fig.\ \ref{Fig4-Cm}(b)]}\footnotetext[5]{Ref.{\onlinecite{09-MS-TbCo2B2C}. The
specific heat of TbCo}$_{{2}}${B}$_{{2}}${C from the present batch\ peaks at
}$T_{p}=5.6(3)$ K{\ }}\label{TabI-Tc-Mag}%
\end{table*}%

The success of these theoretical approaches is impressive, considering that
the calculations of Rhee \textit{et al.} do not admit pairing interactions,
while those for the effective microscopic model have so far been carried out
only in 1D. In both cases, however, theoretical results are only available for
the ground state ($T=0$), thus not covering the rich features of thermodynamic
properties which, for $R(T_{x}M_{1-x})_{2}$\textrm{B}$_{2}$\textrm{C} ($T,M$ =
$3d$ atoms), have been extensively studied\ (see, e.g.,
Refs.\ {\onlinecite{Muller01-interplay-review,Gupta06-Review-Borocarbides}},
and references therein). Evidently, an investigation of the above mentioned
influence of the spacer on both the thermodynamic properties and magnetic
structures would be a welcome contribution for the understanding of the
mechanisms behind many features occurring in the above layered systems (e.g.,
superconductivity in cuprates, heavy fermions, and pnictides materials;
coexistence of superconductivity and magnetic order in magnetic
superconductors, giant magnetoresistance, etc.). As far as the borocarbides
are concerned, such investigation would motivate and guide further theoretical
analyses, in particular those contemplating a unified description of the
observed magnetic and superconducting properties. In pursuit of these
objectives, here we report on the evolution of the magnetic properties of
\textrm{Tb(Co}$_{x}$\textrm{Ni}$_{1-x}$\textrm{)}$_{2}$\textrm{B}$_{2}%
$\textrm{C} ($x=0.2,0.4,0.6,0.8$) when control parameters (such as $x$,
temperature, field, frequency) are varied over a wide measuring range; we use
magnetization, specific heat, ${ac}$-susceptibility, and resistivity
techniques. Our findings will be discussed in terms of the influence of the
electronic properties of \textrm{(Co}$_{x}$\textrm{Ni}$_{1-x}$\textrm{)}$_{2}%
$\textrm{B}$_{2}$ on the observed behavior.

The layout of the paper is as follows. In Sec.\ \ref{sec:Expt} we outline the
measuring techniques employed, while in Sec.\ \ref{sec:Results} the results
are analyzed. Section \ref{sec:Conc} closes the paper with discussions and conclusions.

\section{Experimental}

\label{sec:Expt}

The polycrystalline samples used in this work were the same as the ones used
previously\cite{Tb(CoNi)2B2C-Nd-2012} for collecting neutron diffractograms.
The only difference is that the diffractograms\ and $ac$ susceptibility have
been collected on powdered samples, while the magnetization and specific heat
had been measured on the same small solid piece. This choice of solid form,
instead of a powder one, is dictated by the requirement of the specific-heat
set-up. But this introduces those well-known strong anisotropic
features;\cite{Cho96-TbNi2B2C-anistropy-WF} as a consequence, the shapes of
the field-dependent curves differ considerably from those obtained on powdered
or single crystal samples; nevertheless, the analysis and interpretation of
these curves are straightforward (see below). It is worth adding that
single-crystal studies on this series, though it is highly desirable, would
not invalidate any of the results obtained in this work.

Magnetization and $dc$ susceptibility ($M$, $\chi_{dc}=M/H$, $2<T< 20$ K,
$H\leq90$ kOe) were measured on an extraction-type magnetometer, while the
specific heats ($C$, 2 $<T<$ 40 K and $H=0,$ 30 kOe) were measured on a
relaxation-type calorimeter. $AC$ susceptibilities ($\chi_{ac}=\partial
M/\partial H$, $f\leq$10 kHz, $2\leq T\leq20$ K, $h_{ac}\leq$10 Oe, $H\leq90$
kOe) were measured on a mutual-induction susceptometer.

Longitudinal magnetoresistivity, $\rho(T,H,I\Vert H)$ of a polycrystalline
$x=0.4$ sample was measured by a conventional, home-made, in-line four-point
magnetoresistometer ($0\leq H\leq50$ kOe, $1.8\leq T\leq300$ K, and $0.1\leq
I\leq1$ mA). The residual resistivity ratio, $RRR\equiv\rho(300\mathrm{K}%
)/\rho(1.8\mathrm{K})$, was found to be $\sim1$; this large value is
attributed to the relatively large temperature-independent scattering, arising
from the random distribution of Co/Ni atoms.

Most measurements, in particular those on the $x=0.4$ sample, exhibit strong
hysteresis effects. In order to avoid any complication arising from the
associated history effect, samples were heated up to 50 K (or even higher
temperatures) between successive measurements. Due to these hysteresis
effects, as well as to the widespread and, sometimes, ill-defined shape of the
peak associated with magnetic order, the identification of a unique transition
point was found to be difficult. Accordingly, two definitions of the critical
temperature were adopted (see Table \ref{TabI-Tc-Mag}): (i) $T_{cr}^{\prime
}(x)$ is defined as the point where $C(T,x)$ starts to increase, or the
entropy, $S(T,x)$, starts to deviate downwards away from the high-$T$
extrapolation, though in some cases $T_{cr}^{\prime}(x)$ is manifested as a
shoulder; (ii) the peak-maximum, $T_{p}(x)$, of $C(T,x)$, $\rho(T,x=0.4)$,
$\chi_{dc}(T,x)$, and $\chi_{ac}(T,x)$. The difference between $T_{cr}%
^{\prime}(x)$ and $T_{p}(x)$ is most probably related to atomic distributions.
At any rate, these identifications would not cause any loss of generality or
modify the conclusions reached.

\section{Results and analyses}

\label{sec:Results}

\subsection{Magnetization}%

\begin{figure}[ptbh]%
\centering
\includegraphics[
height=5.9243cm,
width=8.0309cm
]%
{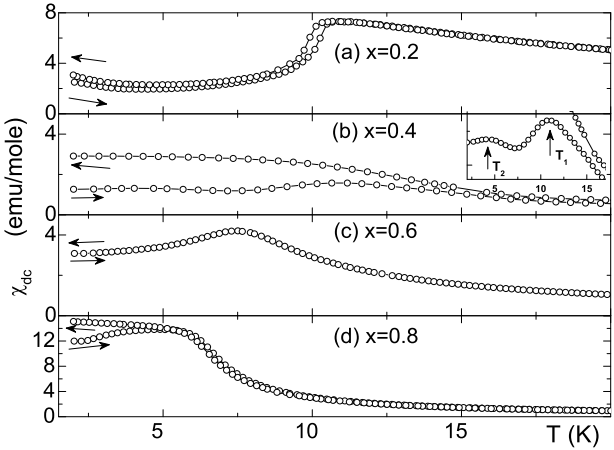}%
\caption{Magnetic susceptibilities\ ($\chi_{dc}=M/H$) for Tb(Co$_{x}$%
Ni$_{1-x}$)$_{2}$B$_{2}$C. For the zero-field curves of $x=0$ and 1, see
Refs.\ \onlinecite{Cho96-TbNi2B2C-anistropy-WF} and
\onlinecite{09-MS-TbCo2B2C}, respectively. Zero-field--cooled and field-cooled
measurements ($H=200$ Oe) were carried out on warming and cooling branch,
respectively. Inset: expansion of the $x=0.4$ curve, showing the $T_{1}$ and
$T_{2}$ events on the warming-up branch (see text). }%
\label{Fig1-dcSus}%
\end{figure}
%

\begin{figure}[ptbh]%
\centering
\includegraphics[
height=9.8782cm,
width=8.0309cm
]%
{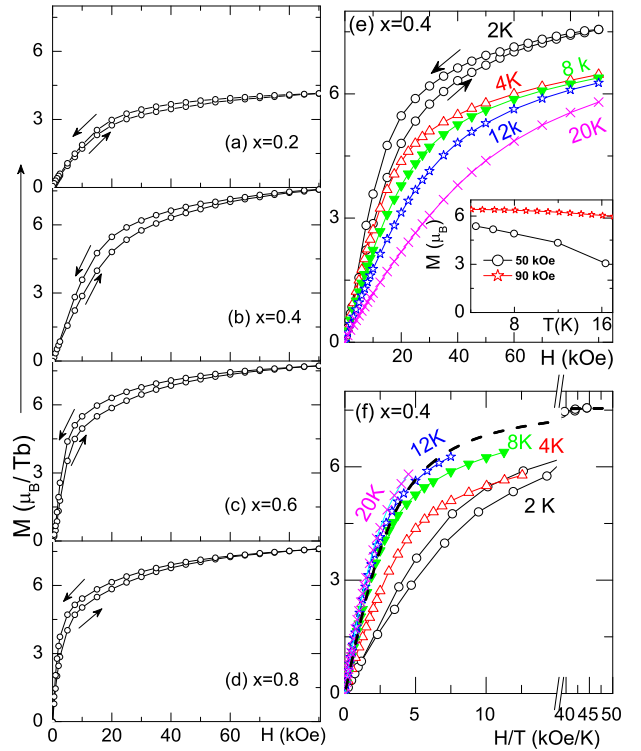}%
\caption{(Color online) Magnetization isotherms of Tb(Co$_{x}$Ni$_{1-x}$%
)$_{2}$B$_{2}$C: (a) $x=0.2$, (b) 0.4, (c) 0.6, and (d) 0.8, measured at $T=2$
K. (e) Magnetization isotherms of the $x=0.4$ sample, measured at 2, 4, 8, 12
and 20K Inset: Thermal evolution of the M($H$, $x=0.4$), $H=$50, 90 kOe,
indicating the absence of any event, neither at $T_{1}$ nor at $T_{1}$.
(f)These same curves are plotted against $H/T$. The dashed line represents the
calculated Brillouin function ($g=1.5$, $J=6$) which is scaled to the
saturation moment of 7.6 $\mu_{B}.$ }%
\label{Fig2-MvsH}%
\end{figure}

Representative $\chi_{dc}(T,x)$ curves are shown in Fig.\ \ref{Fig1-dcSus}.
$\chi_{dc}(T,x=0)$ of \textrm{TbNi}$_{2}$\textrm{B}$_{2}$C\ (not shown)
indicates an AFM-type behavior with a weak FM anomaly at lower
temperatures.\cite{Cho96-TbNi2B2C-anistropy-WF} Similarly, $\chi_{dc}(T,x=1)$
of \textrm{TbCo}$_{2}$\textrm{B}$_{2}$C (not shown) exhibits the onset of the
FM state together with the characteristic ZFC and FC
branches.\cite{09-MS-TbCo2B2C} Some data for these two limiting compositions
have been collected in Table \ref{TabI-Tc-Mag}. $\chi_{dc}(T,x)$ of each
intermediate composition evolves between these two limits. Based on the
magnetic structures identified in Ref.\ \onlinecite{Tb(CoNi)2B2C-Nd-2012}, one
is able to associate the shape of $\chi_{dc}(T,x=0.2)$ with the $\vec{k}_{2}$
AFM mode, and those of $\chi_{dc}(T,x=0.6,0.8)$ with the $\vec{k}_{3}$ and
$\vec{k}_{4}$ modes. By contrast, the shape of $\chi_{dc}(T,x=0.4)$ is
distinctly different, and manifests strong hysteresis effects: during the
warming branch, two events appear (see inset of Fig.\ \ref{Fig1-dcSus}), one
at $T_{1}=11.0\pm0.2$ K, and another at $T_{2}=4.3\pm0.3$ K.%

\begin{figure}[ptbh]%
\centering
\includegraphics[
height=10.2868cm,
width=8.099cm
]%
{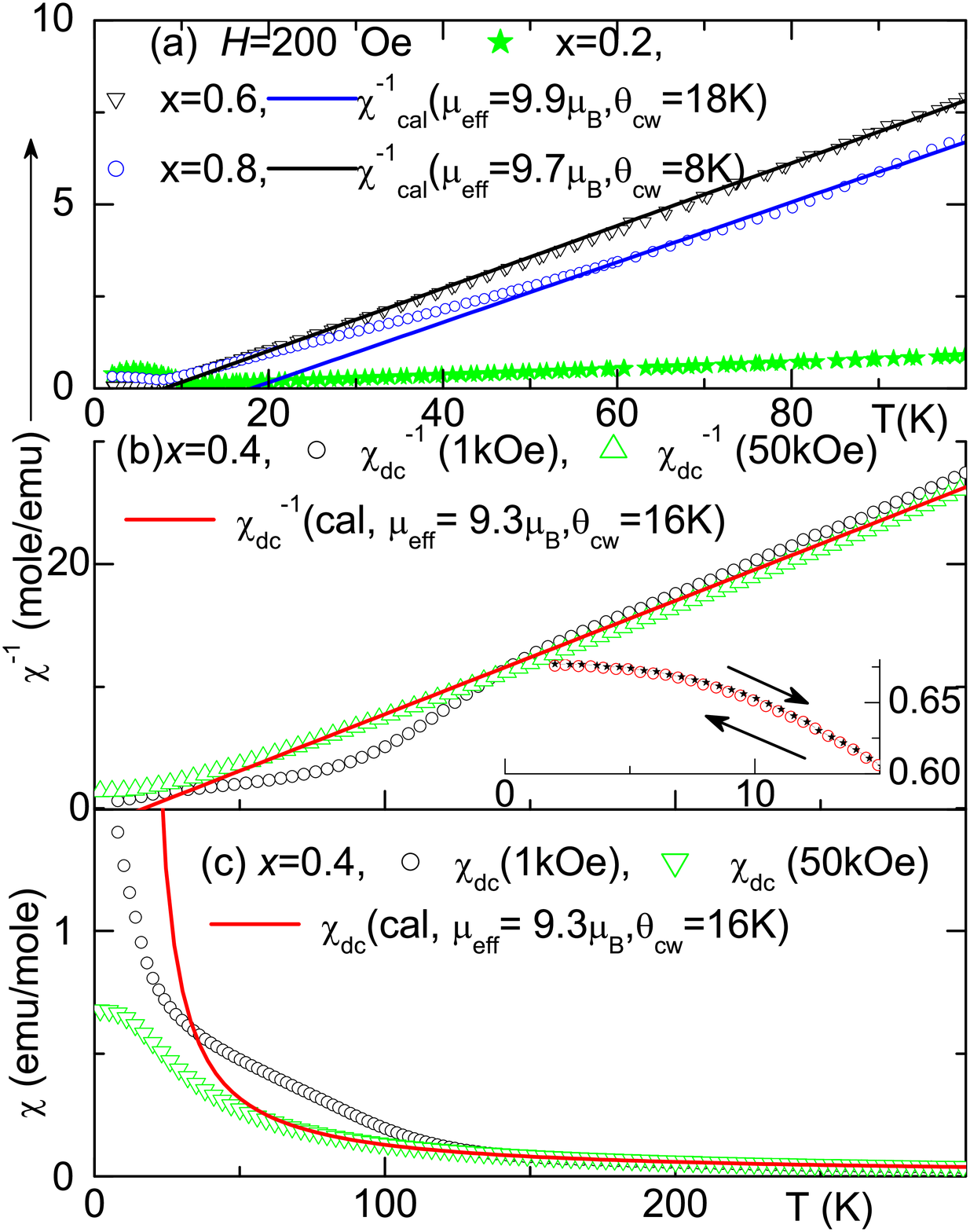}%
\caption{(Color online) (a) Thermal evolution of the inverse $dc$
susceptibility of Tb(Co$_{x}$Ni$_{1-x}$)$_{2}$B$_{2}$C, $x=0.2,0.6,$ and 0.8
for $H=200$ Oe; the solid lines represent the reciprocal of $\chi
_{dc}(T)=0.125\,\mu_{\mathrm{eff}}^{2}/\left(  T-\theta_{CW}\right)  $
emu/mole where $\mu_{\mathrm{eff}}$ (in $\mu_{B}$) is the effective moment,
while $\theta_{CW}$ (in K) is the Curie-Weiss temperature. The deviation of
the $x=0.2$ curve is related to the preferred orientation (see text). (b) Same
as (a), but for the $x=0.4$ sample, and for both $H=1$ kOe and 50 kOe. (c)
Same data as (b), but now is the susceptibility vs. temperature.
\textit{Inset}: $\chi_{dc}(T,x=0.4,H=50$ kOe$)$ on warming (stars) and cooling
(circles) branch, indicating the reduction of hysteresis effects.}%
\label{Fig3-CW}%
\end{figure}
%

\begin{figure}[ptbh]%
\centering
\includegraphics[
height=8.1912cm,
width=8.0309cm
]%
{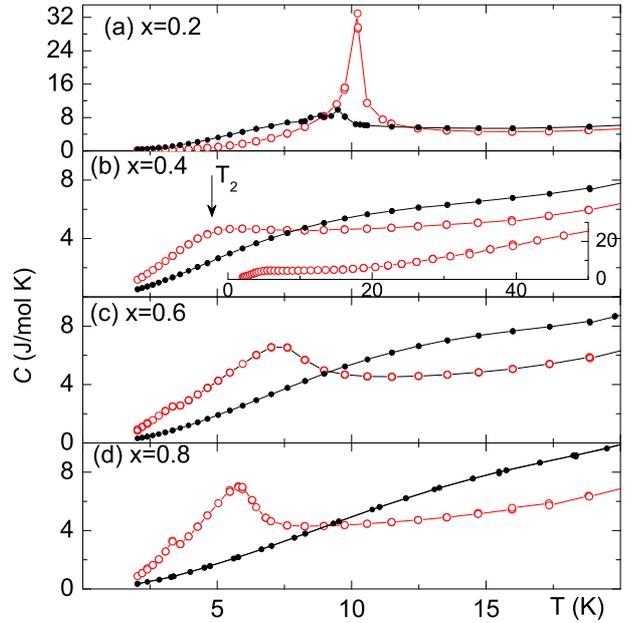}%
\caption{(Color online) Specific heat of Tb(Co$_{x}$Ni$_{1-x}$)$_{2}$B$_{2}$C.
The scales of the vertical axes are not the same. \textit{Open symbols}:
measured at $H=0$; \textit{closed symbols}: measured at $H=30$ kOe.
\textit{Inset}: $C(T,H=0,x=0.4)$ over a wider temperature range. In panel (b),
while the wide shoulder can be attributed to $T_{2}$ event, there are no
visible manifestation of the $T_{1}$ event. The weak shoulder appearing at
$\sim4$ for the $x$=0.6 and 0.8 curves is attributed to the same anomalous
feature reported earlier in TbCo$_{2}$B$_{2}$C (see Ref.
{\onlinecite{09-MS-TbCo2B2C}}).}%
\label{Fig4-Cm}%
\end{figure}
%

\begin{figure}[ptbh]%
\centering
\includegraphics[
height=8.0836cm,
width=8.0309cm
]%
{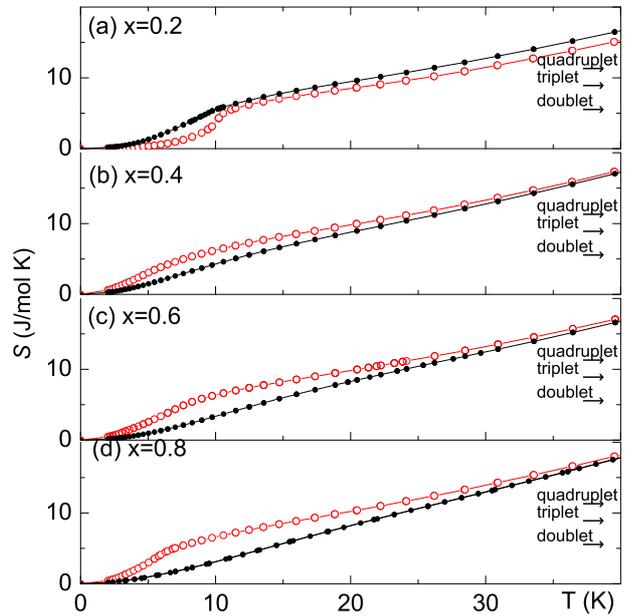}%
\caption{(Color online) Thermal evolution of the calculated entropy of
Tb(Co$_{x}$Ni$_{1-x}$)$_{2}$B$_{2}$C. The horizontal arrows indicate the
entropy associated with a doublet, triplet, and quadruplet state.}%
\label{Fig5-Entropy}%
\end{figure}

Typical $M(2$ K$,H,x)$ isotherms are shown in Fig.\ \ref{Fig2-MvsH}(a)-(e); in
the low-field regime, these confirm the AFM-type character of the $x\leq0.4$
samples, and the FM behavior of the $x\geq0.8$ compositions. In the
$H\rightarrow$ 90 kOe range, the saturated feature of the magnetic moments is
clearly manifested, and the evaluated values are shown in Table
\ref{TabI-Tc-Mag}. In comparison with $\vec{\mu}_{\text{ND}}$, $\left\vert
\vec{\mu}\right\vert _{\text{M(90 kOe)}}(x)$ for $x=0.0,0.6,0.8$ and $1.0$ are
in reasonable agreement, but those for $x=0.2$ and 0.4 differ significantly.
For $x=0.2$, the anomalously lower value of $\left\vert \vec{\mu}\right\vert
_{\text{M(90 kOe)}}$ is attributed to the strong anisotropic character of this
sample, which hinders the saturation to full moment value; indeed, full
saturation is recovered when the magnetization is measured on a powdered
$x=0.2$\ sample. On the other hand, high field data for $x=0.4$ indicate an
almost fully saturated parallel component. This suggests that the process
leading to a reduction in the zero-field $\vec{\mu}_{\text{ND}}$ is no longer
effective upon application of a 90 kOe field; more on this below. In order to
follow this process, we have measured various $M(T,H,x=0.4)$ curves; see Fig.
\ref{Fig2-MvsH}(e). When these are plotted as functions of $H/T$
[Fig.\ \ref{Fig2-MvsH}(f)], three different regimes can be identified. At high
temperatures, all curves collapse very closely onto the Brillouin function,
consistently with a paramagnetic state. Interestingly, upon lowering the
temperature this behavior changes over near $T_{1}$, when the curves start
deviating downwards away from the Brillouin function, signaling the appearance
of a non-zero Weiss molecular field. This persists down to $T\approx T_{2}$,
below which $M(H/T)$ shows appreciable hysteresis, and saturation is reached
at higher $H/T$ values. Two interesting examples of the thermal evolution of
the isofield magnetization are shown in the inset of Fig.\ \ref{Fig2-MvsH}(e):
when fields of 50 and 90 kOe are applied, the moment is almost saturated, and
its thermal evolution shows a smooth monotonic reduction, with no sign of
anomaly at or below $T_{1}$ or $T_{2}$; manifestations of these events only
appear at lower fields.

The high-temperature susceptibility for the $x=0.4,\ 0.6,$ and 0.8 samples
show a well defined Curie-Weiss (CW) behavior [Figs.\ref{Fig3-CW}(a) and (b)],
with $\mu_{\text{eff}}(x=0.4)=9.3\,\mu_{B}$$,\mu_{\text{eff}}(x=0.6)=9.9\,\mu
_{B}$, and $\mu_{\text{eff}}(x=0.8)=9.8\,\mu_{B}$; these are very close to the
theoretical value of $9.72\,\mu_{B}$ for the free Tb$^{3+}$ ion. By contrast,
no such agreement is observed for $x=0.2$: this is related to the strong
preferred orientation [see the discussion of Fig.\ \ref{Fig2-MvsH}(a)].
Nonetheless, it is important to notice from Figs.\ \ref{Fig3-CW}(b) and (c)
that, below 150 K, $\chi_{dc}(T,x=0.4,1\,\mathrm{kOe})$ deviates sharply from
the high-temperature CW behavior, with strong hysteresis effects developing at
lower temperatures, as displayed in Fig.\ \ref{Fig1-dcSus}(b); upon
application of a 50 kOe field, the CW behavior of $\chi_{dc}(T,x=0.4,50$kOe$)$
is observed to survive at much lower temperatures [Figs.\ \ref{Fig3-CW}(b) and
(c)], and no hysteresis effects are manifested [inset of \ref{Fig3-CW}(b)].

\subsection{Specific Heat}

Specific heat curves, $C(T,x)$, are shown in Fig.\ \ref{Fig4-Cm}, while the
calculated entropy, $S(T,x)$ is shown in Fig.\ \ref{Fig5-Entropy}. At such
temperatures, the magnetic (not the weak electronic and lattice) contribution
is dominant. The overall features of these $C(T,x,H=0$) curves depend strongly
on the underlying crystal and magnetic structures. For instance,
Fig.\ \ref{Fig4-Cm} indicates that the $x=0.2$ displays a sharp peak near
$T_{p}(x=0.2)=10$\thinspace K, in marked contrast to all other concentrations;
this is attributed to the influence of a strong magneto-elastic coupling,
which causes the magnetic phase transition to be accompanied by an
orthorhombic structural
distortion.\cite{07-TbNi2B2C,Song99-Tb-magnetostriction,Song01-Tb-highResol-Mag-Xray}%

The thermal evolutions of $C(T,30$ kOe$)$ for all compositions (except for
$x=0.2$) are similar and featureless. Based on the magnetic phase diagrams of
\textrm{TbCo}$_{2}$\textrm{B}$_{2}$\textrm{C} (Ref.\thinspace
\onlinecite{09-MS-TbCo2B2C}) and \textrm{TbNi}$_{2}$\textrm{B}$_{2}$\textrm{C}
(Ref.\thinspace\onlinecite{Cho96-TbNi2B2C-anistropy-WF}), a field of 30 kOe
applied along the $a$ axis would force the Tb moments towards saturation.
Consequently, the magnetic contribution to the low-$T$ (i.e., $T<T_{cr}$)
specific heat of each composition is due to spin-wave excitations from similar
saturated FM states. On the other hand, for $T>T_{cr}$ the contributions are
due to excitations from similar CEF-split levels. The weak peak at $T\approx9$
K for{\textsl{ $x=0.2$}} suggests that for this particular solid sample (with
strongly preferred orientation), the saturation field is higher than 30 kOe;
this feature is consistent with the above-mentioned anisotropy-limited
saturation of the magnetic moment [see Fig.\ \ref{Fig2-MvsH}(a)] and the
anomalous CW behavior of Fig.\ \ref{Fig3-CW}(a).%

\begin{figure}[ptbh]%
\centering
\includegraphics[
height=5.7705cm,
width=8.099cm
]%
{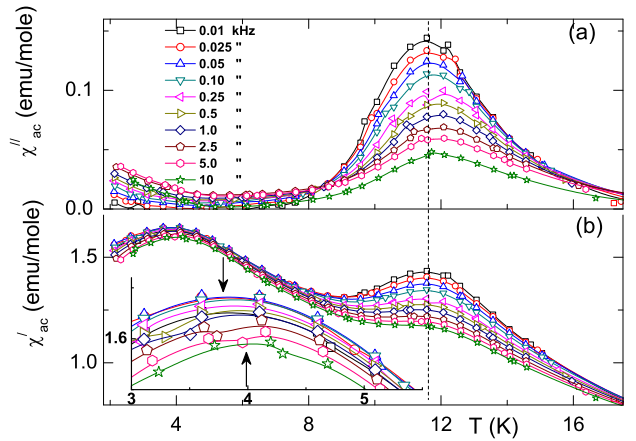}%
\caption{(Color online) Thermal evolution of the $ac$ susceptibility for the
$x=0.4$ sample, taken at $h_{ac}=$10 Oe, and various frequencies.
(a)\ $\chi_{ac}^{\prime\prime}$: the out-of-phase component; (b) $\chi
_{ac}^{\prime}$: the in-phase component. The vertical dashed line emphasizes
the absence of\ a frequency-dependent\ variation in the peak position (though
an intensity reduction is evident). Inset: an expansion of the lower peaks
showing a 10\% reduction in the peak position for four decade frequency
variation (denoted by the short vertical arrows).}%
\label{Fig6-Freq-ACsus-0.4}%
\end{figure}

Figure \ref{Fig5-Entropy} indicates that the thermal evolution of
$S(T<T_{cr},x\geq0.4,H=30\,\mathrm{kOe})$ is very different from that of
$S(T<T_{cr},\ x<0.4,H=30\,\mathrm{kOe})$: for $x>0.4$, the applied field
reduces the thermally-induced randomness in the FM states, as well as in the
$c$-axis modulated mode; similarly, for $x=0.4$ the magnetic disorder is
greatly reduced (see below). By contrast, for $x<0.4$, the disorder of the
AFM-type structure is increased when a smaller $H$ is applied.

Figure \ref{Fig5-Entropy} also shows that $S(T,x,30\,\mathrm{kOe})$ increases
smoothly with $T$, reaching the calculated contributions of a doublet,
triplet, or a quadruplet state and afterwards reaches 17---18 J/moleK at 40 K.
It is noteworthy that: (i) $S(40$ K$)$ is almost the same for all
concentrations, and independent of $H$, indicating that the entropy is
predominantly due to the single-ion character of Tb$^{3+}$; and (ii) a slight
discrepancy between\ $S(40$ K,0 kOe$)$ and $S(40$ K, 30 kOe$)$ is present
solely for $x=0.2$, which is related to the \textit{zero-field}
tetragonal-to-orthorhombic distortion occurring at this concentration; a 30
kOe field suppresses such a distortion and, as a consequence, eliminates the
corresponding contribution.

Finally, $C(T,x=0.4)$ of Fig.\ \ref{Fig4-Cm}(b) displays a shoulder which,
upon comparison with Fig.\ \ref{Fig1-dcSus}(b), can be related to the $T_{2}$
event. By contrast, no sign or anomaly, which can be associated with the
$T_{1}$ event, is exhibited; this suggests that the amount of the involved
entropy is so small that no perceptible change is evident.

\subsection{AC susceptibility}%

\begin{figure}[ptbh]%
\centering
\includegraphics[
height=6.5855cm,
width=8.0309cm
]%
{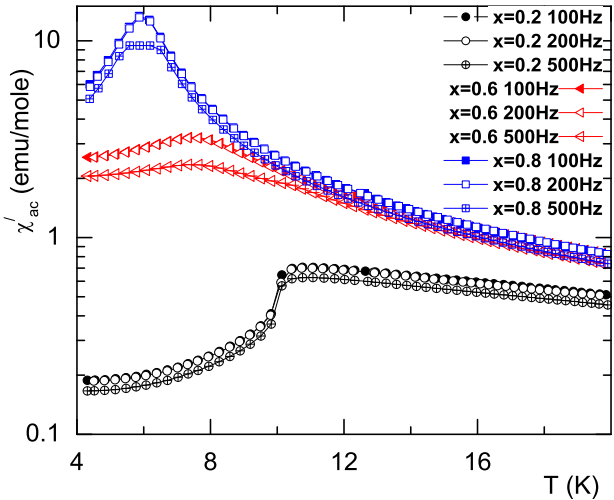}%
\caption{(Color online) Thermal evolution of the zero-\textit{dc}-field, real
component of the \textit{ac} susceptibility of the \textit{x}=0.2,0.6, and 0.8
samples, taken at 100, 200, and 500 Hz, with \textit{h}$_{ac}$=3 to 5 Oe.}%
\label{Fig7-Freq-ACsus-others}%
\end{figure}
%

\begin{figure}[ptbh]%
\centering
\includegraphics[
height=6.3592cm,
width=8.0309cm
]%
{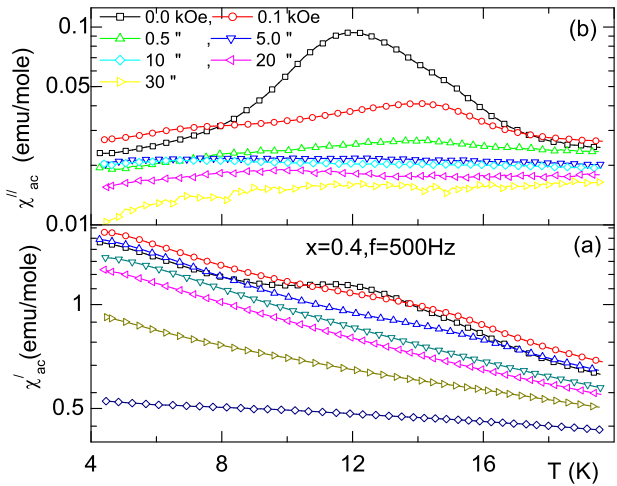}%
\caption{ (Color online) Thermal evolution of the isofield \textit{ac}
susceptibility of \textit{x}= 0.4 sample, taken at 500 Hz. The amplitude of
the oscilating field is \textit{h}$_{ac}$=3$\sim$5 Oe while the range of the
constant applied DC field $H_{dc}$ is between 0 to 30 kOe. (a)\ $\chi
_{dc}^{\prime\prime}$: the out-of-phase component; (b) $\chi_{dc}^{\prime}$:
the in-phase component. }%
\label{Fig8-Field-ACsus-0.4}%
\end{figure}

The discussion so far has unveiled a rather subtle behavior of the $x=0.4$
sample, hence deserving a more thorough analysis. In order to explore whether
there are any dynamic features accompanying its anomalous magnetic behavior,
we have measured its $ac$ susceptibility as a function of frequency $(f)$,
$H$, and $T$: these are shown in Figs.\ \ref{Fig6-Freq-ACsus-0.4} to
\ref{Fig8-Field-ACsus-0.4}. Indeed there are strong and unique dynamic
signatures which are fully emphasized when compared to the conventional
\textit{ac} susceptibility curves of the $x\neq0.4$ samples, shown in
Fig.\ \ref{Fig7-Freq-ACsus-others}. Various points are worth noting. First,
Fig.\ \ref{Fig6-Freq-ACsus-0.4} exhibits two broad peaks, similar to the ones
observed in Fig.\ \ref{Fig1-dcSus}(b), though now they are located at
$T=11.8(2)$ K, and 4.0(2) K; the fact that the peak locations are so close to
the $T_{1}$ and $T_{2}$ events, signaled in Fig.\ \ref{Fig1-dcSus}(b), can
hardly be regarded as fortuitous. Second, the absence of a frequency-dependent
shift at $T_{1}$ indicates that this event is related to a transition from
paramagnetism to long-range order. A closer look at the $\chi_{ac}^{\prime
}(T,$ $x\neq0.4)$ curves in Fig.\ \ref{Fig7-Freq-ACsus-others} shows that the
peak positions do not depend on the frequency, and that they occur at the same
temperature as the maxima of $C(T,x\neq0.4)$ and $M(T,x\neq0.4)$ (see Table
\ref{TabI-Tc-Mag}); this adds credence to the association of $T_{1}$ with a
magnetic transition. Third, the $T_{2}$ event only appears for the $x=0.4$
sample; further, the neutron diffractogram
collected\cite{Tb(CoNi)2B2C-Nd-2012} at $2\,\mathrm{K}<T_{2}$ indicates an
ordered $\vec{k}_{2}$ mode. Therefore, the picture that emerges is that in
this case the system suffers yet another transition at $T_{2}$, occurring
within the magnetically ordered phase.

Finally, Fig.\ \ref{Fig8-Field-ACsus-0.4} shows the influence of a non-zero
\textit{dc} field on the $x=0.4$ sample: we see that both $\chi_{ac}^{\prime}$
and $\chi_{ac}^{\prime\prime}$ are strongly affected even by a weak field,
particularly near $T_{1}$, when the asymmetric peak is strongly reduced, and a
large portion of its intensity weight is shifted to higher temperatures as $H$
is increased. This shift to higher temperature is yet another indication that
ferromagnetic bonds are present in larger fractions, thus playing an important
role at this Co concentration; one can therefore elect these couplings as the
main mechanism behind the $T_{1}$ event.

\subsection{Magnetoresistivity}

\begin{figure}[ptbh]%
\centering
\includegraphics[
height=5.7112cm,
width=8.0309cm
]%
{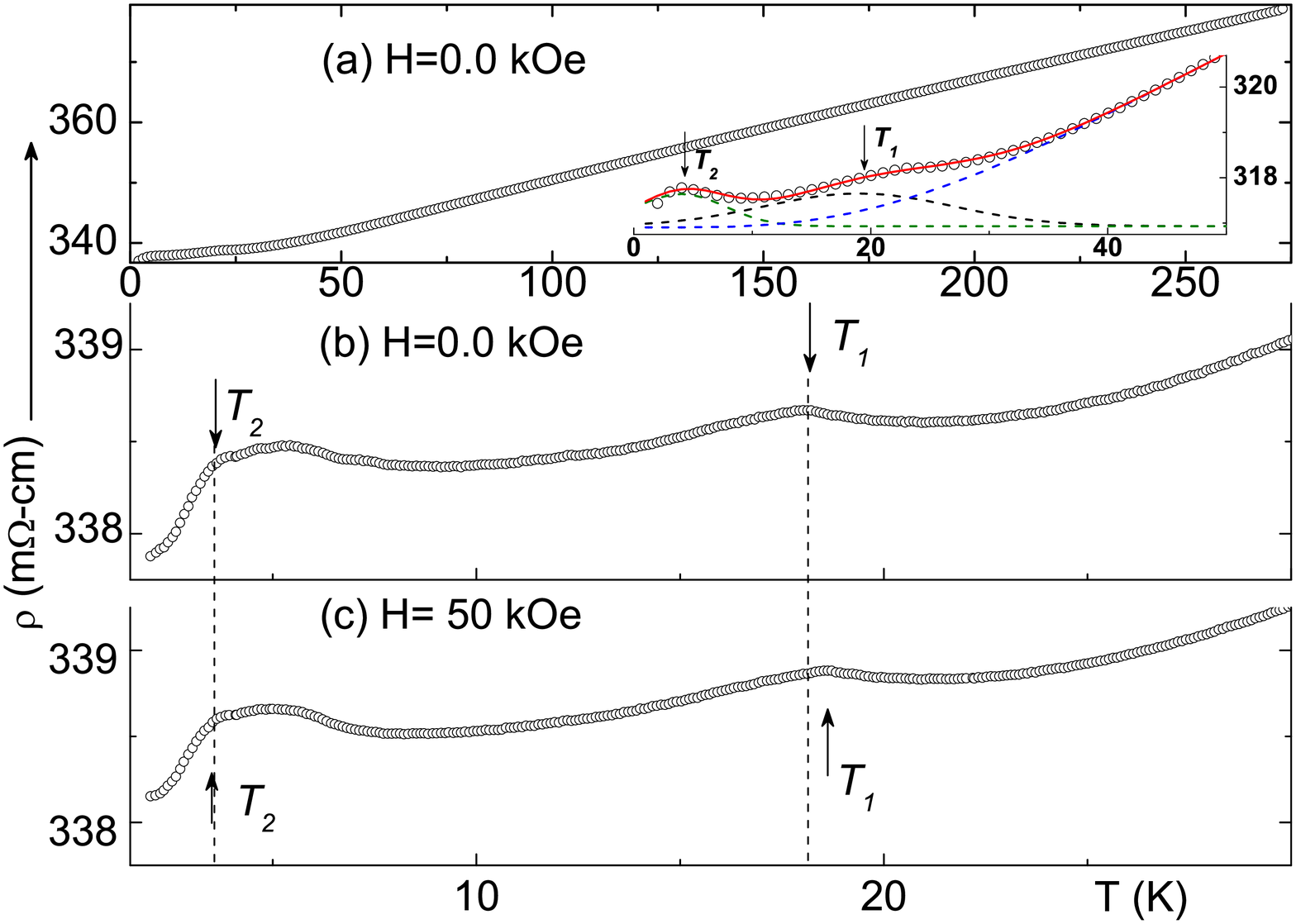}%
\caption{(Color online) (a) Thermal evolution of the zero-field resistivity
for the $x=0.4$ sample. \textit{Inset}: An example of decomposition of the
low-temperature, zero-field resistivity curve into Gaussian contributions
centered at $T_{1}$ and $T_{2}$, and a conventional metallic
Debye--Gr\"{u}neisen
contribution;\cite{Allen96-Quantum-Theory-of-Real-Materials} dashed lines
represent these individual contributions, whose sum is shown as a (red) solid
line. (b) Same as (a), but on an expanded scale. (c) Same as (b), but measured
at 50 kOe. Vertical arrows denote the maxima of the $T_{1}$ and $T_{2}$ peaks,
while dashed vertical lines compare $T_{1}$ and $T_{2}$ measured at $H=0$ and
$50$ kOe. }%
\label{Fig9-RvsTx=0.4}%
\end{figure}

\begin{figure}[ptbh]%
\centering
\includegraphics[
height=6.0166cm,
width=8.0309cm
]%
{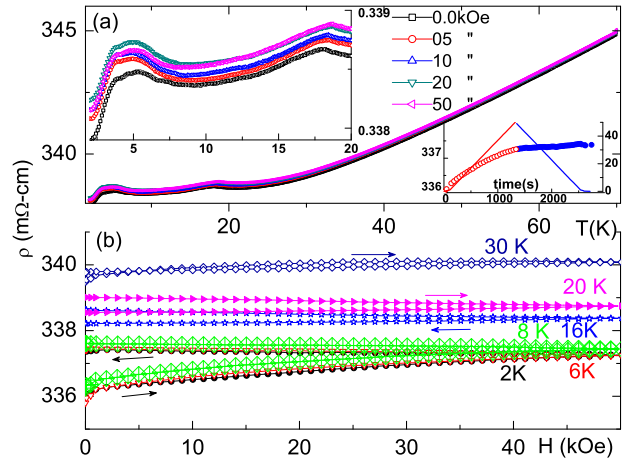}%
\caption{(Color online) (a) Thermal evolution of isofield $\rho(T,H,x=0.4)$
curves measured at various\ fields. \textit{Upper left inset}: expansion of
the low-temperature curves; \textit{lower bottom inset}: \ $\rho(T$=2
K$,H,x=0.4)$ (\textit{symbols}) and applied field (\textit{solid lines}) as a
function of time showing the resistivity evolution on the increasing
(\textit{open symbols}) and decreasing (\textit{solid symbols}) branch of the
applied field. (b) Field-dependent $\rho(T,H,x=0.4)$ isotherms measured at
$T=$ 2, 6, 8, 16, 20, 30 K. The arrows denote the increasing or decreasing
branch of $H$.}%
\label{Fig10-RvsHx=0.4}%
\end{figure}

In many cases, a reduction of $\left\vert \vec{\mu}\right\vert $ may be
attributed to Kondo screening. Though it is not common that $\left\vert
\vec{\mu}\right\vert $ of Tb$^{3+}$ is reduced by such effect, we explored
this possibility by checking whether there is a Kondo-type resistivity minimum
for the $x=0.4$ sample. Representative resistivity curves, $\rho(T,H,x=0.4)$,
are shown in Figs.\ \ref{Fig9-RvsTx=0.4} and \ref{Fig10-RvsHx=0.4}. The
high-$T$ $\rho(T,H=0,x=0.4)$ curve [Fig. \ref{Fig9-RvsTx=0.4}(a)] exhibits a
metallic behavior with a linearity range that extends down to almost 50 K,
below which a typical positive curvature sets in. Figures \ref{Fig9-RvsTx=0.4}%
(b) and \ref{Fig9-RvsTx=0.4}(c) indicate that at low temperatures the
resistivity is dominated by the two events associated with the Gaussian peaks
of the inset of Fig.\ \ref{Fig9-RvsTx=0.4}(a). The intensity, shape, and
position of each peak were found to depend on sample history, similar to what
had already been noted for the magnetization isotherms. Based on our analysis
of the $\chi_{dc}(T)$ and $\chi_{ac}(T)$ curves, these peaks are related to
the $T_{1}$ and $T_{2}$ events, and, further, the observed shift in the peak
positions (as compared to previously identified positions) is attributed to
the different sensitivity of the resistivity probe.%

%

The field dependence of $\rho(T,H,x=0.4)$ is best seen in
Fig.\ \ref{Fig10-RvsHx=0.4}(a). Both the upper inset of
Fig.\ \ref{Fig10-RvsHx=0.4}(a), as well as Fig.\ \ref{Fig10-RvsHx=0.4}(b)
indicate that the hysteresis effect is sharply decreased with temperature, in
particular above $T_{1}$. Moreover, the lower inset of
Fig.\ \ref{Fig10-RvsHx=0.4}(a) indicates that upon increasing $H$, $\rho($2
K$,H,x=0.4)$ increases but surprisingly upon decreasing $H$, no corresponding
decrease is observed in $\rho(T,H,x=0.4)$: this stands in sharp contrast to
the expectation based on the features of Fig.\ \ref{Fig2-MvsH}(a); it suggests
a field-induced metastable state with a longer relaxation time.

The observation that higher fields increase $T_{1}$ is an indication that the
dominant couplings at $T_{1}$ are ferromagnetic; this confirms the findings
drawn from the $\chi_{dc}(T,H,x=0.4)$ and $\chi_{ac}(T,f,H,x=0.4)$ studies.
The intensity of the $T_{1}$ peak in some measurements is relatively strong,
and, as a consequence, the total isofield resistivity curve exhibits a minimum
situated well within the paramagnetic regime. Evidently this is not a Kondo
minimum. Rather, based on the analysis of $\chi_{ac}(T,f,H,x=0.4)$, this is
due to scattering from fluctuation processes accompanying the $T_{1}$ and
$T_{2}$ events.

\section{Discussion and conclusions}

\label{sec:Conc}

It is worth noting that in spite of the above-mentioned layered geometry of
the borocarbides crystal structure, numerous
theoretical\cite{Pickett94-electronic-structure,Matthias94-electronic-structure,Ravindran98-Y(NiCo)2B2C,Lee94-electronic-structure,Coehoorn94-RNi2B2C-electronic-structure}
and
experimental\cite{Cava94-sup-RNi2B2C,Eisaki94-RNi2B2C,Schmidt94-pres-RNi2B2C}
investigations emphasize the \textit{three}-dimensional character of their
electronic structure, transport, and magnetic properties. This is certainly
the case for the present \textrm{Tb(Co}$_{x}$\textrm{Ni}$_{1-x}$%
\textrm{)}$_{2}$\textrm{B}$_{2}$\textrm{C} compositions; accordingly, this
work considers the involved magnetic couplings to be isotropic.

A variation in the ratio of Ni/Co concentrations modifies the Fermi momentum
which in turn alters the spatial scale of the RKKY oscillations, and, as a
consequence, the magnetic couplings; the overall effect is that an
$x$-dependent control of the magnetic properties is achieved.

Further insight into the evolution of these couplings can be gained if we
separate the following three doping regimes: (i) In the $x=0$ limit, the
magnetic structure is a LSDW
mode,\cite{Kawano08-TbNi2B2C-WF,Dervenagas95-mag-struc-TbNi2B2C,Dervenagas96-mag-struct-TbNi2B2C}
resulting from dominant AFM couplings. On the other hand, its high-temperature
CW behavior suggests the presence of sub-dominant FM
couplings\cite{Eisaki94-RNi2B2C} (similar features are evident in
Fig.\ref{Fig3-CW}); nonetheless, these cannot be attributed simply to an
intralayer FM coupling, given the nature of the ensuing LSDW
structure.\cite{Kawano08-TbNi2B2C-WF,Dervenagas95-mag-struc-TbNi2B2C,Dervenagas96-mag-struct-TbNi2B2C}
(ii) For intermediate compositions, $0<x<1$, the dominant couplings for
$x\leq0.4$ are AFM, while they become FM for $x>0.6$. (iii) In the $x=1$
limit, the dominant couplings are FM.\cite{09-MS-TbCo2B2C} The picture that
emerges from these observations is that both FM and AFM couplings are present
in the whole $0\leq x\leq1$ range; however, the relative importance of these
couplings vary monotonically with $x$, being dominated by AFM bonds in the
low-$x$ limit while by FM bonds near the $x\rightarrow1$ region.

This picture allows one to understand the non-linear evolution of $T_{cr}(x)$:
for $x<0.4$, the strength of the dominant AFM\ couplings are progressively
reduced leading to a decrease in $T_{N}(x)$ with $\partial T_{N}/\partial
x\simeq-18$~K$/x$. On the other hand, for $x>0.4$, the magnetic couplings are
also reduced but, being predominantly FM, the rate of reduction of $T_{C}(x)$,
$\partial T_{C}/\partial x\simeq-3$ K$/x$, is almost one sixth of the value
found for the $x<0.4$ samples.

For the particular $x=0.4$ case, both ferromagnetic and antiferromagnetic
couplings are assumed to be present with almost equal strength. Furthermore,
their competitive and opposing tendencies give rise to two magnetic events
which are manifested as (i) a disorder-to-order transition at $T_{1}$, and
(ii) an order-to-order transformation at $T_{2}$. Experimental manifestation
of the $T_{2}$ process is evident in all measurements: as a peak in $\chi
_{dc}(T)$, $\chi_{ac}(T,f,H)$, and $\rho(T,H)$, as a surge of a $\vec{k}_{2}$
mode in neutron diffraction, and as a shoulder in $C(T)$. By contrast, since
the event at $T_{1}$ is assumed to be accompanied by a weak energy of
transformation, then its observation depends on the probing technique;
specifically, it is evident in $\chi_{dc}(T)$, $\chi_{ac}(T,f,H)$, and
$\rho(T,H)$ but, depending on the involved frequency window, it appears at
different temperatures.

Finally, for $x=0.4$, the moment reduction (as seen by the neutron diffraction
below $T_{2}$) and the strong low-temperature deviation from the CW
extrapolation (as seen by $\chi_{dc}(T)$ below 150 K) are unexpected. It is
worth recalling that (i) an application of a magnetic field leads to a full
moment saturation at 2 K, accompanied by a metastable state, (ii) the
evolution of the 50 and 90 kOe induced magnetic moment is smooth, and with no
sign of anomalies, at or below $T_{1}$ and $T_{2}$, and that (iii) the
Brillouin function is almost followed for $T>T_{1}$. A Kondo screening, or
magnetic fluctuations, could in principle provide possible mechanisms for the
moment reduction; however, our resistivity results exclude any influence of a
Kondo screening, while the observation of the $T_{1}$ event in $\chi_{dc}(T)$
rules out any significant role played by magnetic fluctuations. Spin glass
features\cite{Mydosh96-SpinGlass} could also be invoked, given that $\chi
_{ac}(T,f)$ exhibits a small, but noticeable frequency-dependent
shift-in-temperature (see the inset of Fig.\ \ref{Fig6-Freq-ACsus-0.4});
however, these should also be discarded, since neutron diffraction indicates
well ordered magnetic and crystalline structures. The observation that CW
behavior is recovered above 150 K is a strong indication that the above
mentioned anomalies are related to CEF configurations. In the absence of an
applied magnetic field, they modify the level scheme of the free Tb$^{3+}$ ion
in such a way that the zero-field moment is strongly reduced, the CW behavior
is disrupted below 150 K, and strong hysteresis and relaxation effects are
induced. An applied magnetic field re-arranges the CEF-split levels in such a
way that the moment is restored to the values observed in neighboring
compositions, the CW behavior survives to much lower temperatures, and
hysteresis effects are reduced.

In summary, the $x$-dependent evolution of the thermodynamic properties of
\textrm{Tb(Co}$_{x}$\textrm{Ni}$_{1-x}$\textrm{)}$_{2}$\textrm{B}$_{2}%
$\textrm{C} was investigated by a wide range of experimental techniques. The
magnetic properties across the whole concentration range have been found to be
consistent with the findings of the neutron diffraction studies: in
particular, the dominant magnetic couplings for $x\leq0.4$ composition are
found to be AFM (in agreement with the observed AFM modes), while those, for
$x>0.6$, are strongly FM (also in agreement with the observed FM modes).
Furthermore, the magnetic transition temperatures, as well as the Tb magnetic
moments are observed to evolve non-monotonically with the Ni/Co ratio. These
unexpected features are highly emphasized in the anomalous behavior of the
$x=0.4$ sample, which, in addition, exhibits a zero-field moment reduction and
a surge of two magnetic events (argued to be related to disorder-to-order and
an order-to-order transitions). These features are attributed to combined
influences of competing magnetic couplings and crystalline electric field effects.

\begin{acknowledgments}
Partial financial support from the Brazilian Agencies CNPq, CAPES, and FAPERJ
is gratefully acknowledged.
\end{acknowledgments}

\bibliographystyle{apsrev}
\bibliography{borocarbides,crystalography,intermetallic,interplay-sup-mag,MagClassic,massalami,ND-RepAnalysis,notes,To-Be-Published}

\end{document}